\documentclass[prl,showpacs,preprint,endfloats]{revtex4}

\newcommand {\etal}{{\it et al.}}

\usepackage[dvipdfmx]{graphicx}
\usepackage{bm}
\usepackage{pifont}
\usepackage{amsmath,amssymb}
\usepackage[dvipdfmx]{color} 

\begin{document}

\title{Magnetic field-induced insulator-semimetal transition in a pyrochlore Nd$_2$Ir$_2$O$_7$}

\author{K. Ueda$^1$}
\author{J. Fujioka$^{1}$}
\author{B.-J. Yang$^{2}$}
\author{J. Shiogai$^{3}$}
\author{A. Tsukazaki$^{3}$}
\author{S. Nakamura$^{3}$}
\author{S. Awaji$^{3}$}
\author{N. Nagaosa$^{1,2}$}
\author{Y. Tokura$^{1,2}$}

\affiliation{
$^1$ Department of Applied Physics and Quantum Phase Electronics Center, University of Tokyo, Tokyo 113-8656, Japan\\
$^2$ Center for Emergent Matter Science (CEMS), RIKEN Advanced Science Institute (ASI), Wako 351-0198, Japan\\
$^3$ Institute for Materials Research, Tohoku University, Sendai 980-8577, Japan}

\date{March 1, 2015}

\begin{abstract}
We have investigated magneto-transport properties in a single crystal of pyrochore-type Nd$_2$Ir$_2$O$_7$.
The metallic conduction is observed on the antiferromagnetic domain walls of the all-in all-out type Ir-5$d$ moment ordered insulating bulk state, that can be finely controlled by external magnetic field along [111].
On the other hand, an applied field along [001] induces the bulk phase transition from insulator to semimetal as a consequence of the field-induced modification of Nd-4$f$ and Ir-5$d$ moment configurations.
A theoretical calculation consistently describing the experimentally observed features suggests a variety of exotic topological states as functions of electron correlation and Ir-5$d$ moment orders which can be finely tuned by choice of rare-earth ion and by magnetic field, respectively. 
\end{abstract}
\pacs{71.30.+h, 72.80.Ga, 78.20.-e, 78.30.-j}
\maketitle

Topological electronic states in quantum materials have been subjects of intensive experimental and theoretical studies in condensed matter research\cite{rev-Hasan,2012A3Bi,2014Na3Bi,Cd3As2ARPES,NcommYang,2015Na3Bi}.
Recently, cooperative effects of electron correlation and spin-orbit coupling have attracted much attention as the central ingredient to synthesize the next-generation topological quantum phase\cite{Krempa_rev}.
A Weyl semimetal (WSM) state is one representative example where linearly dispersing bands with the band-crossing Weyl point (WP) show up in a three-dimensional bulk while hosting surface Fermi arc states\cite{2007Murakami,XWan}, which is proposed to be realized in various correlated magnetic materials\cite{XWan,HgCr2Se4,HgCr2Se4-PGS,CaOs2O4,2012Krempa,PRBGangChen,2013PRLSBLee,PRXYamaji,PRLYang}.

Pyrochlore-type $R_{2}$Ir$_{2}$O$_{7}$ ($R$: rare-earth, Y, and Bi ion) is one of the promising candidates for the realization of abundant emergent phases including WSM\cite{XWan,2012Krempa,PRBGangChen,2013PRLSBLee,PRXYamaji,PRLYang,NPhysBalents,PRBYang,2011Imada}.
The pyrochlore lattice consists of the nested corner-sharing tetrahedra of each $R$ ion and Ir one, as displaced from each other by a half unit cell (Fig. 1 (a))\cite{pyrochlore-rev}.
Owing to large spin-orbit coupling, both $R$-4$f$ and Ir-5$d$ spins possess a single-ion magnetic anisotropy denoted by broken lines in the lower panel of Fig. 1 (a), which leads to various magnetic ground states\cite{1997Harris,1998Bramwell,1998Harris,spinice,2002Fukazawa}.
For instance, the all-in all-out (AIAO)-type configuration may be favored with the nearest-neighbor antiferromagnetic exchange coupling where all of four spins on the vertices of the pictured tetrahedron point in (4-in 0-out, 4/0) or out of (0-in 4-out, 0/4) its center\cite{1998Bramwell}.
Since this ordering pattern breaks time-reversal symmetry while retaining cubic symmetry, WSM state is predicted to exist with WP at Fermi energy in some parameter regions\cite{2012Krempa,PRBGangChen,PRLYang}.
On the other hand, when we apply a magnetic field along [001] direction ($H||[001]$) and the Zeeman energy overcomes the exchange coupling, two spins point inwards and other two outwards, forming the so-called 2-in 2-out (2/2) state\cite{2002Fukazawa,PRLTaguchi,2007Machida}.
Such a change of magnetic structure may reconstruct band structures, which can produce unconventional magneto-transport phenomena\cite{2013PRLSBLee,2011PIO}.

The pyrochlore with $R=$ Nd offers an ideal platform to explore the evolution of electronic state with the change in magnetic configuration where not only the magnetic moment of Ir-5$d$ electrons but also Nd-4$f$ local moment plays a crucial role\cite{PRBGangChen,PRLYang}.
When it undergoes a metal-insulator transition accompanied by the AIAO-type Ir-5$d$ magnetic ordering, Nd-4$f$ moments also develop the AIAO-type ordering at lower temperature due to $f$-$d$ exchange coupling\cite{2011Matsuhira,NIO-neutron,NIO-MIT,EIO-Xray,NIO-muon}, although there are few conclusive evidences about the ordering patterns or different conclusions\cite{Disseler_muon}.
Conversely, we may have a chance of largely modifying the Ir-5$d$ spin ordering pattern in terms of magnetic-field control of Nd-4$f$ moments via the $f$-$d$ exchange coupling.
In fact, characteristic magneto-transport phenomena such as large magnetoresistance are observed in polycrystals, although these origins remain elusive\cite{2013Disseler,NIO-GMR}.
Another remarkable feature for Nd$_2$Ir$_2$O$_7$ is an anomalous metallic state on the AIAO-type magnetic domain walls (DWs), contrary to the gapped bulk state\cite{NIO-MIT,NIO-DW}.
These facts indicate that the electronic state strongly couples to the magnetic features in this material, and the study on field-direction dependent magnetotransport in $R=$ Nd single crystals is indispensable.

In this Letter, we have investigated the magneto-transport properties with a field-tuned magnetic structure in a $R=$ Nd single crystal.
AIAO-type magnetic DWs hosting metallic states, which clearly show up in the hysteresis both in resistivity and magnetization, can be finely controlled by $H||[111]$.
In contrast, the insulating state turns into a semimetallic one under $H||[001]$ where the magnetic structure of Nd-$4f$ moment is in the 2-in 2-out configuration.
Combined with our mean-field calculation, we suggest that the observed phase transition can be attributed to the topologically-nontrivial band reconstruction with the variation of magnetic structures.

\begin{figure}
\begin{center}
\includegraphics[width=15cm]{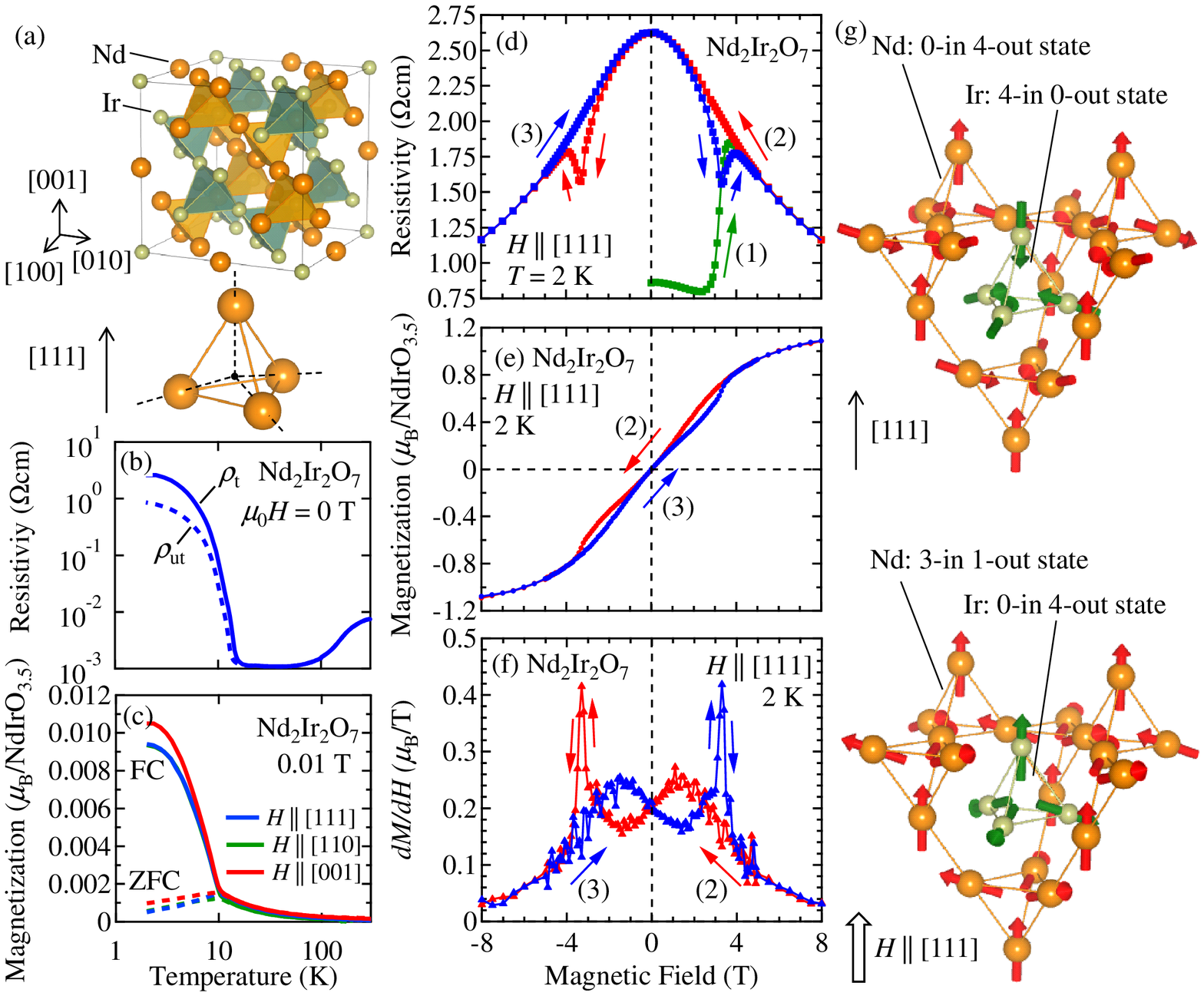}
\caption{(color online) 
(a) Schematic of the Nd$_2$Ir$_2$O$_7$ unit cell and one tetrahedron composed of four ions.
The broken lines in the tetrahedron denote the magnetic easy axes. 
Temperature dependence of (b) resistivity of untrained state $\rho _{\rm{ut}}$ and that of trained state $\rho _{\rm{t}}$, and (c) magnetization in zero field cooling (ZFC) and field cooling (FC). 
Magnetic field dependence of (d) resistivity, (e) magnetization, and (f) field derivative of magnetization for $H|| [111]$. The green curve denotes the virgin curve, the red one denotes the sweeping process (2), and the blue one denotes the process (3), respectively.
(g) Schematic of the possible magnetic structures in Nd-4$f$ moments (red arrows) and Ir-5$d$ moments (green arrows). Top figure shows the configuration in low field, and the bottom figure shows it in sufficiently strong field along $[111]$. 
}
\end{center}
\end{figure}

High-quality single-crystalline samples of $R=$ Nd were synthesized by KF flux method described in Ref.\cite{Millican}.
We polished the present crystal into a well-formed bar-shape. Its dimension is about $0.75\times 0.65\times 0.30$ mm, suitable for precise transport measurement.
A standard four-probe method was employed for resistivity measurements with an electric current along [1$\bar{1}$0] crystalline axis perpendicular to an applied magnetic field.
The resistivity is measured down to 2 K and up to 14 T by using Physical Property Measurement System (Quantum Design).
AC resistance measurement up to 25 T at 27 mK was performed at 1 kHz with excitation current of 100 $\rm{\mu}$A by using a dilution refrigerator embedded in the cryogen-free hybrid magnet at High Field Laboratory for Superconducting Materials in Institute of Materials Research, Tohoku University\cite{dilution_imr}.

We show the temperature dependence of resistivity and magnetization in Fig. 1 (b) and (c), respectively.
The resistivity in zero field sharply increases with decreasing temperature below 15 K ($=T_{\rm{N}}$), at which the magnetization in zero-field-cooling (ZFC) starts to deviate from that in field-cooling (FC) pointing to the onset of the AIAO-type magnetic ordering of Ir-5$d$ moment.
These anomalies are similar to the data for polycrystalline samples and can be attributed to the AIAO-type magnetic ordering of Ir-$5d$ moment\cite{2011Matsuhira,NIO-MIT}.
The resistivity in the domain-aligned state (termed trained state, $\rho _{\rm{t}}$) is larger than that in the multi-domain state (untrained state, $\rho _{\rm{ut}}$), indicating the presence of the metallic DW in the multi-domain state, as observed in the polycrystals\cite{NIO-DW}.

To quantify the relation between the alignment of magnetic domain and resistivity in more detail, we show the magnetic field dependence of resistivity and magnetization at 2 K under $H||[111]$ in Figs. 1 (d)-(f).
The initial value of resistivity is as small as $\sim 0.8$ $\rm{\Omega cm}$ and shows irreversible jump around $+3.4$ T on process (1) in Fig. 1 (d) due to the field-elimination of DWs.
With decreasing field from $+14$ T (process (2)), the resistivity monotonically increases, reaches the maximum around 0 T, and then the resistivity decreases with a characteristic dip around $-3.4$ T.
The resistivity with increasing field (process (3)) shows a similar profile to that in process (2), with a dip around $+3.4$ T.
Such hysteresis of magnetoresistivity may be attributed to the transient multi-domain state upon the magnetic domain flipping between 4/0 and 0/4 configuration driven by $H||[111]$ as predicted in Ref. {\cite{JPSJArima}} and experimentally demonstrated in Ref. \cite{Tardif}.
Another sign of the domain flip is observed in the magnetization as shown in Fig. 1 (e).
It monotonically increases as a function of field and saturates in a high magnetic field around $\sim 1.2$ $\mu _{\rm{B}}\rm{/NdIrO_{3.5}}$.
Since the magnitude of magnetization is nearly governed by the Nd-$4f$ moment ($\sim 2.37$ $\mu _{\rm{B}}$/mol\cite{NIO-neutron}) rather than Ir-$5d$ moment ($\sim 0.2$ $\mu _{\rm{B}}$/mol\cite{NIO-neutron,YIO-neutron}), it is anticipated that the magnetic structure of Nd-sublattice gradually turns into three-in one-out (3/1) configuration from 0/4 one at zero field so as to reconcile with the Zeeman energy, as frequently observed in pyrochlore oxides\cite{2002Fukazawa,PRLTaguchi,2007Machida}.
We note that the magnetization shows a small hysteresis below 4 T and tiny kinks at $\pm 3.4$ T, as also reported in Ref.\cite{2013Disseler,NIO-DW}.
These are more clearly identified in the profile of  $dM/dH$ shown in Fig. 1 (f).
It clearly exhibits a peak at $+3.4$ T ($-3.4$ T) in process (2) (process (3)), which corresponds to the kink in magnetization.
These observed properties provide insights about the magnetic structure under $H||[111]$ as depicted in Fig. 1 (g).
In low fields, the Nd-$4f$ moment forms 0/4 configuration stabilizing the Ir-$5d$ 4/0 state as depicted in the top figure of Fig. 1 (g).
As the field increases, the Nd-$4f$ moment may gradually turn into 3/1 state due to the competition between Zeeman energy and exchange coupling. Such modulations of Nd-sublattice can flip the 4/0 magnetic domain of Ir-sublattice to 0/4 one through the $f$-$d$ exchange coupling as shown in the bottom of Fig. 1 (g).

\begin{figure}
\begin{center}
\includegraphics[width=15cm]{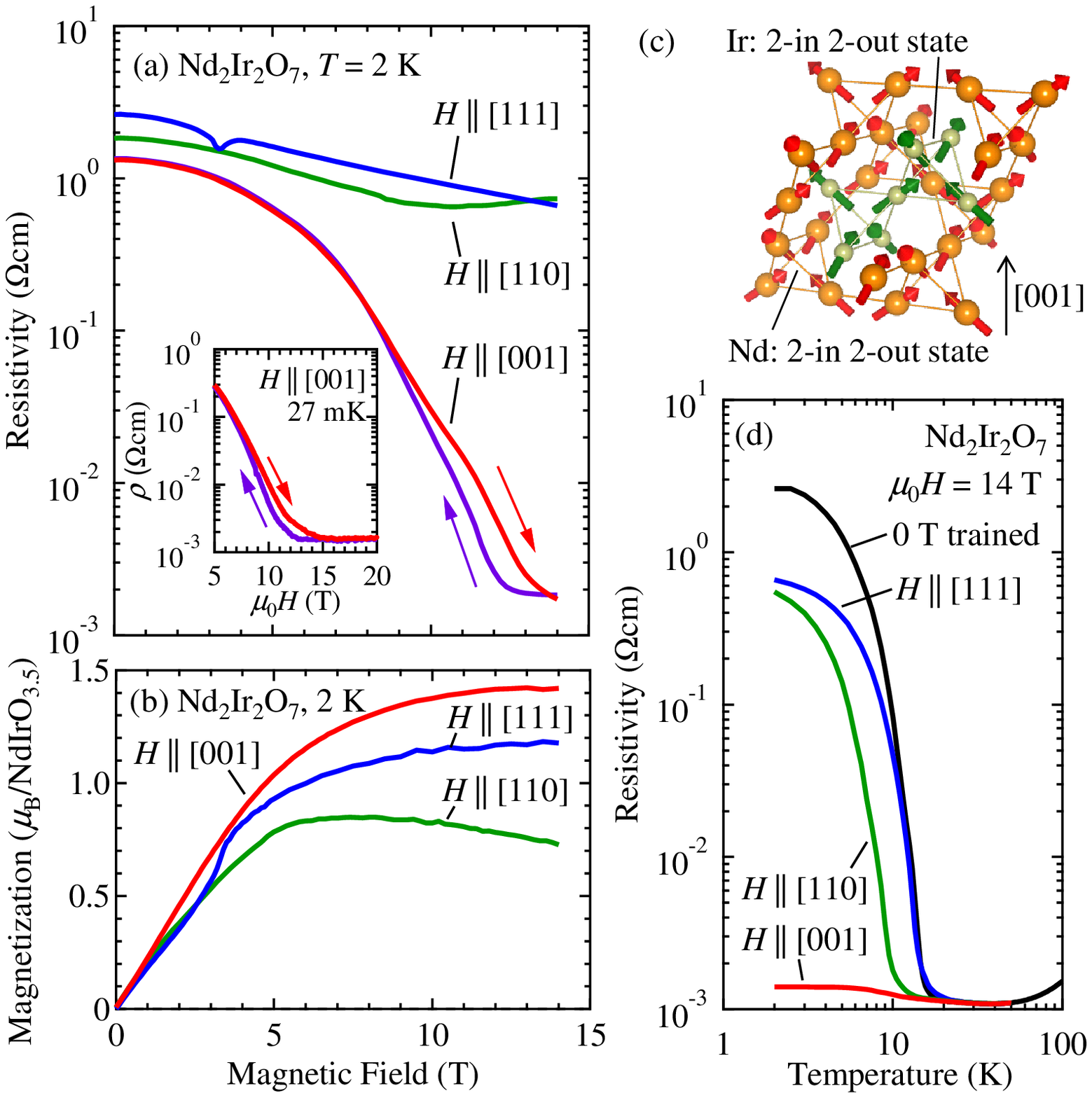}
\caption{(color online). 
Magnetic field dependence of (a) resistivity and (b) magnetization at 2 K for several field directions. The blue curve denotes $H||[111]$, the green one $H||[110]$, and the red (purple) one $H||[001]$ on increasing (decreasing) field process, respectively. The inset to (a) shows the field dependence of resistivity for $H|[001]|$ up to 20 T at 27 mK.
(c) Schematic of the possible spin configurations in $H||[001]$.
(d) Temperature dependence of $\rho _{\rm{t}}$ (black line), $\rho (\rm{14 T})$ along $H||[111]$ (blue line), $H||[110]$ (green line), and $H||[001]$ (red line), respectively.
}
\end{center}
\end{figure}

Figure 2 shows the field dependence of resistivity and magnetization at 2 K for $H||[111], [110],$ and $[001]$, respectively.
The negative magnetoresistance is observed for all field directions\cite{H110}.
Remarkably, the resistivity for $H||[001]$ decreases by three orders of magnitude at 14 T and shows a clear hysteresis between increasing and decreasing field process above 8 T.
The inset of Fig. 2 (a) shows the result of the $H||[001]$ scan up to above 20 T at 27 mK.
The resistivity saturates and reaches the minimum of $\sim 1.6$ $\rm{m\Omega cm}$ closing the hysteresis curve above 15 T, reminiscent of the first-order insulator-semimetal transition induced by a magnetic field in colossal magnetoresistance manganites\cite{science_Kuwahara}. 
The magnetization for $H||[001]$ monotonically increases as field increases, saturating in high field up to $\sim 1.4$ $\mu \rm{_{B}/NdIrO_{3.5}}$, ensuring the formation of the Nd-4$f$ moment 2/2 configuration\cite{2002Fukazawa,PRLTaguchi,2007Machida}.
The 2/2-type Nd-4$f$ configuration may turn the AIAO-type Ir-5$d$ configuration into the 2/2-type one via $f$-$d$ exchange interaction as illustrated in Fig. 2 (c); this is likely the origin of the observed insulator-semimetal transition.
Moreover, as shown in Fig. 2 (d), the field along [001] direction nearly smears out the upturn of resistivity upon the semimetal-insulator transition with lowering temperature, whereas the resistivity increases rapidly below $T_{\rm{N}}$ in other field directions.

\begin{figure}
\begin{center}
\includegraphics[width=15cm]{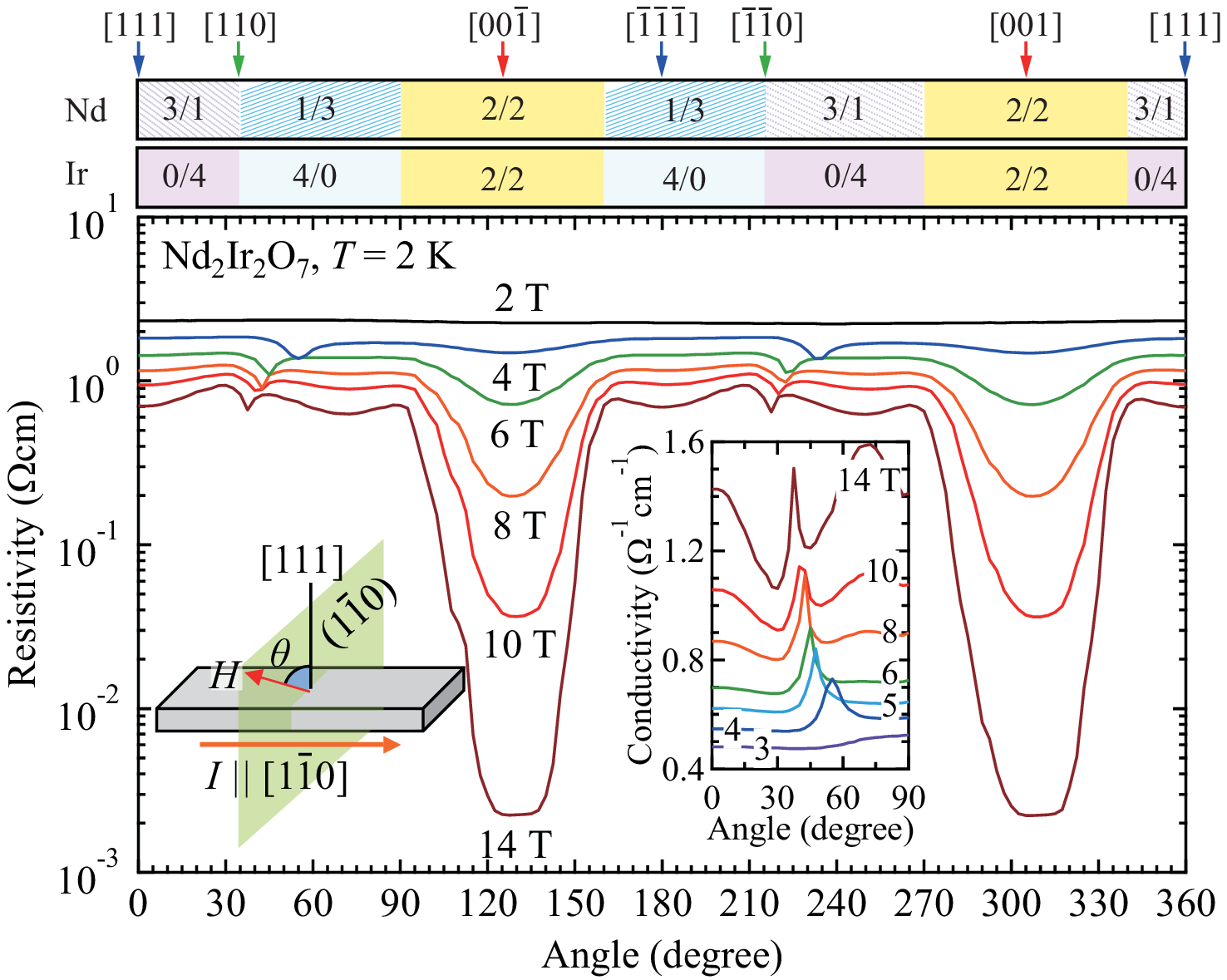}
\caption{(color online). 
The two top tables denote the possible magnetic structures of Nd-4$f$ and Ir-5$d$ moments in strong field limit. The numbers denote the magnetic moment configurations as described in the text. 
The bottom figure shows the angle dependence of resistivity (on a logarithmic scale) for several fields. The left inset exhibits the experimental configuration. The right inset shows the low angle dependence of conductivity (on a linear scale) for several fields. 
}
\end{center}
\end{figure}

To confirm the electronic band structure reconstruction driven by the change in magnetic orders, we have investigated the field-angular dependence of the resistivity in Fig. 3.
Prior to the measurements, we apply $H|[111]$ to realize the 4/0 single-domain state.
The origin of rotation angle ($\theta = 0^{\circ }$) is defined to be $H||[111]$ direction, and the field is rotated around [1$\bar{1}$0] axis parallel to the current direction (see the lower left inset of Fig. 3), passing through along the $[110]$-axis at $\theta = 35^{\circ }$, the $[00\bar{1}]$-axis at $\theta = 125^{\circ }$, and finally reaches the $[\bar{1}\bar{1}\bar{1}]$-axis at $\theta = 180^{\circ }$.
The right lower inset of Fig. 3 shows the magnified figure of conductivity (inverse of resistivity) on a linear scale.
The peak structure appears right above $\theta = 35^{\circ }$ at 14 T, shifts to a larger angle position with decreasing field, and suddenly disappears below 3 T.
The magnitude of conductivity peak observed above 4T is almost independent of the field amplitude.
Moreover, around $\theta = 125^{\circ }$, significant reductions of resistivity are observed and its magnitude strongly depends on the strength of magnetic field.
Similar behavior is identified in the region of  $180^{\circ }$ - $360^{\circ }$, suggesting the $180^{\circ }$ periodicity.
The conductivity peak (resistivity dip) structure around $\theta = 35^{\circ }$ is clearly attributed to the flipping of Ir-5$d$ moment between 4/0 and 0/4 configuration which accompanies the proliferation of metallic DW, since $\theta = 35^{\circ }$ is the boundary between the different Nd-4$f$ moment textures of 1/3 and 3/1 patterns in the high-field limit as shown in the top label of Fig. 3.
Another remarkable feature is the reduction of resistivity in the range of $90^{\circ }$ - $160^{\circ }$, where the 2/2 state can be favored in the strong field limit.
Such a sudden decrease of resistivity cannot be explained in terms of conventional angle dependent resistivity in the fixed band structure nor of scattering by magnetic impurities, rather pointing to the field-induced electronic-structure modulation coupled to the 2/2 magnetic structure.

\begin{figure}
\begin{center}
\includegraphics[width=15cm]{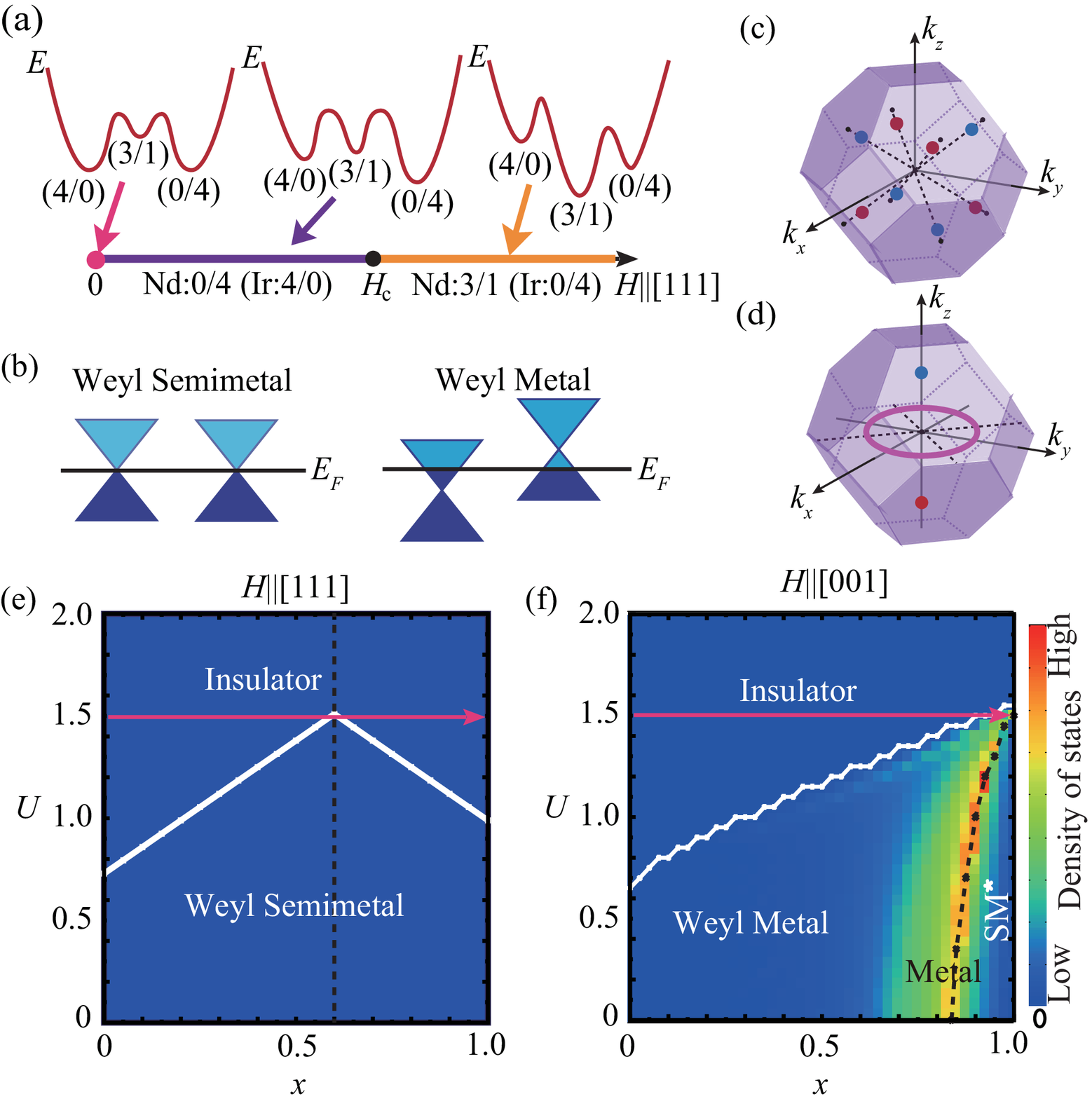}
\caption{(color online).
(a) Schematic of the free energy $E$ as a function of $H||[111]$.
(b) Schematic band structures near Fermi energy $E_{\rm{F}}$ for Weyl semimetal (WSM) and Weyl metal (WM).
Distribution of nodal points in the momentum space for (c) WSM state with AIAO Ir-5$d$ magnetic pattern and (d)nodal semimetal (SM*) with 2/2 one. A red (blue) dot indicates a Weyl point with the chiral charge +1 (-1) and the purple ring denotes a line node.
Electronic phase diagram based on a mean-field theory for (e) $H||[111]$ and (f) $H||[001]$, respectively. The horizontal axis ($x$) denotes the continuous evolution of Nd-4$f$ spin configuration from 0/4 to 3/1 states in (e) and from 0/4 to 2/2 in (f). 
The red arrow indicates the evolution of the system as $H$ increases, which can consistently describe the magneto-transport observed in experiment.
}
\end{center}
\end{figure}

To elucidate the role of $f$-$d$ exchange coupling and electron correlation in $R=$ Nd, we have performed a self-consistent mean-field study of an extended Hubbard model\cite{2012Krempa,PRBGangChen}, the details of which are given in the Supplemental Material\cite{SupplementalMaterial}.
Let us first consider the case of $H||[111]$.
Due to Zeeman energy, Nd-4$f$ configuration evolves from 0/4 to 3/1 as $H$ increases.
Accordingly, the effective magnetic field ($h_{\rm{eff}}$) induced by $f$-$d$ exchange coupling forces the Ir-5$d$ configuration to change from 4/0 to 0/4 as described in Fig. 4 (a).
Since both $U$ and $h_{\rm{eff}}$ favor 4/0 or 0/4 configuration, they work cooperatively\cite{PRBGangChen}.
Namely, the critical value of $U$ ($U_{\rm{c}}$) for the WSM-insulator transition becomes smaller as $h_{\rm{eff}}$ gets larger.
The field-induced variation of $h_{\rm{eff}}$ is described by using a parameter $x\in [0,1]$.
Here $x=0$ ($x=1$) indicates the case when the Nd-sublattice forms 0/4 (3/1) configuration uniformly through the sample in small (large) $H$ limit.
The range of $0<x<1$ describes the continuous evolution of Nd-4$f$ configuration between these two limits ($x=M/M^{\rm{sat}}$).
Since $h_{\rm{eff}}$($x=0$) and $h_{\rm{eff}}$($x=1$) have the opposite sign, it should vanish at a certain point $x_{\rm{c}}\in (0,1)$.
In fact, this is the point ($x_{\rm{c}}=0.6$) where $U_{\rm{c}}$ is at maximum in Fig. 4 (e).
Since both the WSM and the insulator have zero density of states on the Fermi level ($D(E_{\rm{F}})=0$), the magneto-transport would not show an abrupt change as $H$ varies, being consistent with the experiment.
Although Zeeman coupling of Ir-5$d$ electrons can induce small electron-hole pockets in WSM due to broken cubic symmetry, the Zeeman energy scale is so tiny that its contribution to transport is negligible\cite{SupplementalMaterial}.
On the other hand, when $H||[001]$, Nd-sublattice develops the 2/2 configuration to induce the 2/2 Ir-5$d$ state via $f$-$d$ coupling in the large $H$ limit.
It is worth to note that the 2/2 Ir-5$d$ state has a completely different nature as compared to the AIAO one which is basically a WSM having 8 WPs on the Fermi level (Fig. 4 (c)).
When $U<U_{\rm{c}}$, the ground state of 2/2 Ir-5$d$ configuration is a nodal semimetal\cite{HgCr2Se4,HgCr2Se4-PGS,topologicalnodalsemimetal} (SM* in Fig. 4 (f)) with a line node on the $k_{\rm{z}}=0$ plane and two WPs on the $k_{\rm{z}}$ axis (Fig 4 (d)).
Since the states on the line node are not degenerate, small electron/hole pockets appear on the $k_{\rm{z}}=0$ plane.
Similarly, two WPs on the $k_{\rm{z}}$ axis develop small electron pockets.
However, since the size of these electron/hole pockets are so tiny, $D(E_{\rm{F}})$ is still very small.
Such a huge difference in the nature of the Femi surface topology between 2/2 and AIAO Ir-5$d$ states gives rise to intriguing magneto-transport properties of the system.
For small $H$ (or small $x$) where the 4/0 Ir-5$d$ configuration dominates over the 2/2 one, the phase diagram is basically similar to the case of $H||[111]$.
However, since the small 2/2 Ir-5$d$ component breaks the cubic symmetry, the WSM turns into a Weyl metal (WM) with small electron/hole pockets and hence $D(E_{\rm{F}})$ becomes nonzero.
As $H$ (or $x$) grows, we can see a sharp increase of $D(E_{\rm{F}})$ in a small window.
In fact, this is the region where the Fermi surface topology changes.
The distortion of the band structure to mediate two topologically distinct semimetals WM and SM* induces a huge $D(E_{\rm{F}})$ as shown in Fig. 4 (f), which is assigned to the physical origin of the observed metallic behavior when $H||[001]$.
The theoretical calculation with $U_{\rm{c}}\sim 1.5$ (in the unit of electron hopping interaction) appears to well account for the observed properties as shown in the red arrows in Figs. 4 (e) and (f).

In conclusion, we have presented the magneto-transport properties upon the magnetic field-modulation of magnetic structure in a $R=$ Nd single crystal.
$H||[111]$ can align the AIAO-type single-domain state and switch two variants to each other, leading to the characteristic hysteresis in resistivity and magnetization.
In contrast, $H||[001]$ turns the insulating phase with AIAO magnetic structure into the unconventional semimetallic phase with 2/2 state.
The theoretical calculation suggests that  the magnetic field-induced switching of Nd-4$f$ spin configurations from AIAO to 2/2 state modifies the topological nature of the Ir-5$d$ band structure, which can account for the unconventional magneto-transport properties.

We are grateful to T. Kurumaji, Y. Yamaji, and M. Imada for fruitful discussions.
This work is supported by the Japan Society for the Promotion of Science through the Funding Program for World-Leading Innovative R$\& $D on Science and Technology (FIRST Program) on 'Quantum Science on Strong Correlation' initiated by the Council for Science and Technology Policy and by JSPS Grant-in-Aid for Scientific Research (No. 80609488 and No. 24224009).


\newpage

\begin{thebibliography}{100}
\bibitem{H110} The decrease of magnetization is discerned in the high field region for $H||[110]$; this might be due to the strong diamagnetism arising from the interband orbital effect as frequently observed in a narrow-gapped or gapless system, since $H||[110]$ can hardly modify the AIAO configuration of Ir-5$d$ moments.
\end{thebibliography}
\end{document}